\documentclass[12pt,epsf]{article}
\usepackage{amsmath}
\usepackage{amssymb}
\usepackage{bm}
\usepackage[dvips]{graphics}

\newcommand{\be}{\begin{equation}}
\newcommand{\ee}{\end{equation}} 
\newcommand{\bea}{\begin{eqnarray}}
\newcommand{\eea}{\end{eqnarray}}

\newcommand{\al}{\alpha}

\newcommand{\hPhi}{\hat{\Phi}}

\newcommand{\hW}{\hat{W}}

\newcommand{\hV}{\hat{V}}

\newcommand{\vev}[1]{\left\langle{#1}\right\rangle}
\newcommand{\resol}{\frac{1}{z-\hPhi}}

\newcommand{\coeff}{\frac{1}{64\pi^2}}

\newcommand{\hN}{\hat{N}}
\newcommand{\mmfac}{\frac{\hN}{g_m}}
\newcommand{\mmfacinv}{\frac{g_m}{\hN}}

\newcommand{\cF}{{\cal F}}
\newcommand{\cN}{{\cal N}}
\newcommand{\der}{\partial}

\def\tr{\text{tr}}
\def\Tr{\text{Tr}}

\begin{document}
\setlength{\oddsidemargin}{0cm}
\setlength{\baselineskip}{7mm}

\begin{titlepage}  \renewcommand{\thefootnote}{\fnsymbol{footnote}}
$\mbox{ }$
\begin{flushright}
\begin{tabular}{l}
KUNS-1950 \\
RIKEN-TH-34 \\
ROMA1-1398/04 \\
hep-th/0412216\\
December 2004
\end{tabular}
\end{flushright}

~~\\
~~\\
~~\\

\vspace*{0cm}
    \begin{Large}
       \vspace{2cm}
       \begin{center}
         {Direct derivation of the Veneziano-Yankielowicz superpotential \\ from matrix model}
\\
       \end{center}
    \end{Large}

  \vspace{1cm}

\begin{center}

          Hikaru K{\sc awai}$^{ab}$\footnote
           {
E-mail address : hkawai@gauge.scphys.kyoto-u.ac.jp},
          Tsunehide K{\sc uroki}$^b$\footnote
           {
E-mail address : kuroki@riken.jp},
          Takeshi M{\sc orita}$^a$\footnote
           {
E-mail address : takeshi@gauge.scphys.kyoto-u.ac.jp}{\sc and}
          Kensuke Y{\sc oshida}$^c$\footnote
           { 
E-mail address : kensuke.yoshida@roma1.infn.it}           

$^a$           {\it Department of Physics, Kyoto University,
Kyoto 606-8502, Japan}\\
$^b$           {\it Theoretical Physics Laboratory, RIKEN (The
  Institute of Physical and Chemical Research), Wako, Saitama 
351-0198, Japan}\\
$^c$           {\it Dipartimento di Fisica, Universit\`a di Roma ``La Sapienza''\\
P.le Aldo Moro, 2 - 00185 ROMA, Italy\\
and\\
 I.\ N.\ F.\ N., Sezione di Roma I}\\

\end{center}

\vfill

\begin{abstract}
\noindent 
We derive the Veneziano-Yankielowicz superpotential directly 
from the matrix model by fixing the measure precisely. 
The essential requirement here is that the effective superpotential 
of the matrix model corresponding to the ${\cal N}=4$ 
supersymmetric Yang-Mills theory vanishes except 
for the tree gauge kinetic term. Thus we clarify 
the reason why the matrix model reproduces the Veneziano-Yankielowicz 
superpotential correctly in the Dijkgraaf-Vafa theory.

\end{abstract}

\vfill
\end{titlepage}
\vfil\eject

\setcounter{footnote}{0}
\section{Introduction}
\setcounter{equation}{0}

It has been revealed that the connection between gauge theory 
and a matrix model is deep and interesting. In particular, 
the large-$N$ reduced model \cite{Kawai:1982nm} is not only 
useful because it reduces the dynamical degrees of freedom 
and thus makes the large-$N$ gauge theory tractable, 
but it would provide possibly a constructive definition 
for a gauge theory, or even string theory \cite{Ishibashi:1996xs}. 
For $\cN=1$ supersymmetric gauge theory, Dijkgraaf and Vafa 
proposed that a simple matrix model also has enough information 
on the F-term of the effective superpotential 
\cite{Dijkgraaf:2002dh}. More precisely, in the $\cN=1$ supersymmetric 
gauge theory coupled to a chiral superfield $\Phi$ in the adjoint 
representation with a superpotential 
\be
W(\Phi)=\sum_{k=0}^{n}\frac{g_k}{k+1} \Phi^{k+1}, 
\label{potential} 
\ee
the prepotential $\cF (S,g_k)$ 
($S=\frac{1}{64\pi^2}\tr W^{\al}W_{\al}$) is equivalent 
to the free energy $F_m(g_m,g_k)$ of a one-matrix model
\be
S_m=\frac{\hN}{g_m}\Tr W(\Phi),  
\label{DV action}
\ee    
in the large-$\hN$ limit under an identification 
$S=g_m$.\footnote
{Here we have assumed that there is no gauge symmetry breaking.} 

The proofs of their proposal are given in 
\cite{Dijkgraaf:2002xd,Cachazo:2002ry}. In particular, 
it is shown in \cite{Cachazo:2002ry} by using the Konishi anomaly 
\cite{Konishi:1983hf}
that the Schwinger-Dyson equation for 
$\coeff\vev{\tr\left(\frac{W^{\al}W_{\al}}{z-\Phi}\right)}$
is exactly the same as that for the resolvent of the matrix model 
$\mmfacinv\vev{\Tr\left(\resol\right)}$ 
in the large-$\hN$ limit.  Because the former 
and the latter is given by $\der \cF/\der g_k$ and 
$\der F_m/\der g_k$ respectively, we find that the prepotential 
$\cF$ and the free energy of the matrix model $F_m$ are 
equivalent up to a function independent of $g_k$'s. 
However, as noted in 
\cite{Ooguri:2002gx,Dijkgraaf:2002dh,Cachazo:2002ry}, 
the matrix model 
produces a stronger result than the above consideration. 
By taking the superpotential $W=m\Phi^2/2$ and 
under a suitable identification 
between the matrix model measure and the gauge theory cutoff, 
$F_m$ can also reproduce the $g_k$-independent part of $\cF$ 
that corresponds to the Veneziano-Yankielowicz (VY) superpotential 
\cite{Veneziano:1982ah}
\begin{align}
S\left[\log\left(\frac{\Lambda^{3N}}{S^N}\right)+N\right],
\label{VY}
\end{align}
where $\Lambda$ is the dimensional transmutation scale 
associated with the gauge dynamics. 
In this sense, the connection 
between the $\cN=1$ gauge theory and the matrix model 
seems deeper than we have expected. 

In \cite{Kawai:2003yf}, it is shown that the Dijkgraaf-Vafa 
theory can be regarded as the large-$N$ reduction. 
This enables us to construct a direct map between correlators 
in the gauge theory and those in the matrix model and thus 
to show directly that equalities hold between them. 
From this point of view, it must be possible 
to find the origin of the VY superpotential in the matrix model, 
because we have a direct map between the gauge theory and the 
matrix model including the gauge field degrees of freedom.
 
In this paper, we show that the matrix model indeed has an 
information on the pure gauge field degrees of freedom 
and that it can reproduce the VY superpotential. 
In particular, by a matrix model consideration 
we can derive exactly the key identification mentioned above 
between the measure in the matrix model 
and the cutoff in the gauge theory, 
which is just assumed in  \cite{Cachazo:2002ry}. 
Evidently in order to do this, it is indispensable to fix 
the measure in the matrix model. 
We do this by requiring that the free energy of 
the matrix model corresponding to the $\cN=4$ supersymmetric 
gauge theory must vanish except for a term that corresponds to 
the tree gauge kinetic term. 
It is quite natural to fix the $g_k$-independent part of the 
free energy in this way, because it is well-known 
that the $\cN=4$ gauge theory is a finite theory and does not have 
any quantum corrections to the holomorphic part of the effective Lagrangian 
\cite{Buchbinder:2004bu}. 
Then we clarify the reason why the matrix model also reproduces 
the pure gauge contribution to the prepotential from the 
point of view of the large-$N$ reduction.\footnote
{Derivations of the VY superpotential from the field 
theory point of view in the context of the Dijkgraaf-Vafa theory 
are given, for example, in \cite{Gripaios:2003kt,Arnone:2004ek}. 
In the former, it is derived by introducing fundamental matters
to the $\cN=1$ gauge theory, 
while in the latter it is done by invoking the $\cN=4$ theory.} 

In section 2 we determine the measure in the matrix model 
based on the above idea. Using this measure, we derive 
the VY superpotential in section 3. Section 4 is devoted to 
conclusions. In the appendix we present the derivation 
of the VY superpotential in the case of the broken gauge symmetry 
as an application of our approach.  
 
\section{Determination of the measure in the matrix model} 
\setcounter{equation}{0}
In this section we determine the measure in the matrix model 
according to our requirement mentioned in the introduction. 

We begin with an $\cN=1$ supersymmetric $U(N)$ gauge theory 
coupled to three chiral multiplets $\Phi_i$ ($i=1\sim 3$) 
in the adjoint representation with the following potential:\footnote
{We will concentrate on the $SU(N)$ part for simplicity.} 
\begin{align}
S =& \int d^4x d^2 \theta ~ 2\pi i \tau_0 ~ 
\tr \left( W^\alpha W_\alpha \right)
+ \int d^4x d^2 \theta ~ 
\tr \left(  \Phi_1[\Phi_2,\Phi_3] + W(\Phi_1) 
+ \frac{m_2}2 \Phi_2^2 +\frac{m_3}2 \Phi_3^2 \right) \nonumber \\
+& \int d^4x d^2 \theta d^2 \bar{\theta} ~ 
\sum_{i=1}^3 \tr
\left( e^{-V} \bar{\Phi}_i e^{V} \Phi_i \right) + c.c. ,
\label{original}
\end{align}
where $W(\Phi)$ is the superpotential given in \eqref{potential}.
If we take $W(\Phi_1)=m_1\Phi_1^2/2$, that is, 
$g_1=m_1$ and $g_k=0$ for $k\ge 2$, this theory 
is nothing but what is called the $\cN=1^*$ theory, 
which becomes in the limit $m_i \rightarrow 0$ ($i=1\sim 3$)
the $\cN=4$ supersymmetric Yang-Mills theory (SYM) 
in terms of $\cN=1$ superfields. 
On the other hand, if we take $m_2=m_3=\Lambda_0 \gg m_1$, 
at a scale below $\Lambda_0$, $\Phi_2$ and $\Phi_3$ are decoupled  
and the theory becomes the $\cN=1$ supersymmetric 
gauge theory coupled to the chiral multiplet $\Phi_1$ 
with the superpotential $W(\Phi_1)$. 

In Dijkgraaf-Vafa theory, 
as far as the holomorphic part of the effective action 
is concerned, we can drop the kinetic terms for the vector 
and the chiral multiplets and have only to consider a matrix model 
corresponding to the superpotential 
\begin{align}
S_m = \frac{\hat{N}}{g_m} 
\Tr \left( \Phi_1[\Phi_2,\Phi_3] + W(\Phi_1)
+\frac{m_2}2 \Phi_2^2 +\frac{m_3}2 \Phi_3^2 \right), 
\label{1^**}
\end{align}
where $\Phi_i$ ($i=1\sim 3$) is an $\hN\times \hN$ Hermitian 
matrix. The free energy $F_m$ of this matrix model is defined by 
\be
e^{-\frac{\hat{N}^2}{g_m^2}F_m}
=C \int d \Phi_1 d \Phi_2 d \Phi_3 e^{-S_m},
\label{F}  
\ee
where $d\Phi_i$ is the standard measure, and $C$ is an appropriate 
measure factor. 

In order to determine $C$, we consider in particular 
the matrix model corresponding to the $\cN=1^*$ theory 
\begin{align}
S_{\cN=1^*}= \frac{\hat{N}}{g_m} 
\Tr \left( \Phi_1[\Phi_2,\Phi_3] + \frac{m_1}2 \Phi_1^2
+\frac{m_2}2 \Phi_2^2 +\frac{m_3}2 \Phi_3^2 \right). 
\label{1^*}
\end{align}
Its free energy is given as in \eqref{F} by  
\begin{align}
e^{-\frac{\hat{N}^2}{g_m^2}F_{\cN=1^*}}=Z_{\cN=1^*}
=C \int d \Phi_1 d \Phi_2 d \Phi_3 e^{-S_{\cN=1^*}}.  
\label{N=4F}
\end{align}
We then use the fact that the holomorphic part 
of the effective Lagrangian in $\cN=4$ SYM is given simply 
by the tree gauge kinetic term: 
\be
\cF_{\cN=4}=\frac{\pi i\tau_0}{N} S^2,~~~ 
W^{\text{eff}}_{\cN=4}=N\frac{\der \cF_{\cN=4}}{\der S}=2\pi i\tau_0 S,  
\label{gauge kinetic term}
\ee
where $\tau_0$ is the bare coupling constant.   
Identifying $\cF_{\cN=4}$ and $S$ with $ F_{\cN=4}$ 
and $g_m$,\footnote
{The identity $S=g_m$ can be shown directly by using
the map constructed in \cite{Kawai:2003yf}.}
respectively, and using $F_{\cN=4}=\lim_{m_i\rightarrow 0} F_{\cN=1^*}
$
we have the key equation that determines $C$:
\begin{align}
\lim_{m_i\rightarrow 0} F_{\cN=1^*}
=\frac{\pi i\tau_0}{N} g_m^2.
\label{requirement}
\end{align}
 
Integrating out $\Phi_1$ in \eqref{N=4F}, we obtain 
\begin{align}
Z_{\cN=1^*}
&= C \left( \frac{2\pi g_m}{\hat{N}m_1} \right)^{\frac{\hat{N}^2}2} 
\int d\Phi_2 d\Phi_3 e^{-S'},  \\
S'=&\frac{\hat{N}}{g_m} 
\Tr \left( -\frac{1}{2m_1}
[\Phi_2,\Phi_3]^2+\frac{m_2}2 \Phi_2^2 +\frac{m_3}2 \Phi_3^2 \right).
\end{align}
Then we diagonalize $\Phi_2$ and set $(\Phi_2)_{ii}=b_i$. 
The integration over the angular variables of $\Phi_2$ 
gives 
\begin{align}
\int d \Phi_2 = J \int d b_{i} \prod_{i>j}(b_i-b_j)^2,
\label{Jacobian}
\end{align}
where $J$ is a constant determined below, which simply originates 
from the change of the variables, and is independent of the action 
of the matrix model. 

Setting $(\Phi_3)_{ij}=c_{ij}$, the partition function 
can be expressed as 
\begin{align}
Z_{\cN=1^*}
&=C J \left( \frac{2\pi g_m}{\hat{N}m_1} \right)^{\frac{\hat{N}^2}2} 
\int d b_i  d c_{ij}  \prod_{i>j}(b_i-b_j)^2
 e^{-S''}, \nonumber \\
 S''&= \frac{\hat{N}}{g_m}
\left( \sum_{i\neq j} \frac{|c_{ij}|^2}2 \left( \frac{1}{m_1}(b_i-b_j)^2
+m_3 \right)+ \sum_{i} \frac{m_3}2 c_{ii}^2+
\sum_{i} \frac{m_2}2 b_i^2 \right). 
\end{align}
The integration with respect to $c_{ij}$ can be readily performed 
to yield 
\begin{align}
Z_{\cN=1^*}=
C J \left( \frac{2\pi g_m}{\hat{N}} \right)^{\hat{N}^2} 
\left(\frac{1}{m_1 m_3} \right)^{\frac{\hat{N}}2}
\int d b_i   \prod_{i>j}\frac{(b_i-b_j)^2}{ \left((b_i-b_j)^2
+m_1 m_3\right)}
 e^{-\frac{\hat{N}m_2}{2g_m}\sum_{i}b_i^2}.
\end{align}
When $m_1m_3\ll 1$, we find 
\begin{align}
\int d b_i   \prod_{i>j}\frac{(b_i-b_j)^2}{ \left((b_i-b_j)^2
+m_1 m_3\right)}
 e^{-\frac{\hat{N}m_2}{2g_m}\sum_{i}b_i^2}
=& \left( \frac{2\pi g_m}{\hat{N}m_2}\right)^{\frac{\hat{N}}2}, 
\end{align}
therefore, 
\begin{align}
Z_{\cN=1^*}=C J \left( \frac{2\pi g_m}{\hat{N}} \right)^{\hat{N}^2} 
\left(\frac{2\pi g_m}{\hat{N}m_1 m_2 m_3} \right)^{\frac{\hat{N}}2}.
\end{align}
We see that the contributions from the mass terms become 
subleading in the large-$\hN$ limit.  
Therefore, in this limit we can take the $\cN=4$ limit 
$m_i\rightarrow 0$ smoothly and obtain  
\begin{align}
\frac{\hN^2}{g_m^2}\lim_{m_i\rightarrow 0} F_{\cN=1^*}
=-\log\left(C J \left( \frac{2\pi g_m}{\hat{N}} \right)^{\hat{N}^2} 
      \right). 
\end{align} 
{}From our requirement \eqref{requirement}, we can fix the measure factor 
$C$ as
\be
C=J^{-1}\left(\frac{\hN}{2\pi g_m}\right)^{\hN^2}
e^{-\pi i\tau_0 \hN^2/N}. 
\label{C}
\ee 

For the computation of $J$, it is sufficient 
to consider a concrete example, because $J$ is independent 
of the action as mentioned above. 
A convenient choice is the Gaussian action. 
A straightforward integration 
yields  
\begin{align}
Z=\int d \Phi e^{-\frac{1}2 \tr \Phi^2}
=& (2\pi)^{\frac{\hat{N}^2}2},  
\label{gauss}
\end{align}
where $\Phi$ is an $\hN\times \hN$ Hermitian matrix.  
On the other hand, by using \eqref{Jacobian}, we obtain 
\begin{align}
\int d \Phi e^{-\frac{1}2 \tr \Phi^2}
=& J \int d p_i \prod_{i>j}(p_i-p_j)^2 e^{-\frac{1}2 p_i^2}. 
\end{align}
This can be computed by means of the orthogonal polynomials, 
which are defined as 
\begin{align}
\int dx& e^{-\frac{1}{2} x^2} P_n(x) P_m(x) = \delta_{nm} h_n, 
\nonumber \\
P_n(x)&=x^n+\cdots. 
\end{align}
In the case of the Gaussian action,
they are nothing but the Hermite polynomials and we have 
$h_n=n! (2\pi)^{\frac{1}{2}}$. 
Therefore, the partition function can be also expressed as 
\begin{align}
Z=J \hat{N}! \prod_{i=0}^{\hat{N}-1}h_i 
= J \hat{N}! (\hat{N}-1)! \dots 0! (2\pi)^{\frac{\hat{N}}2}.
\end{align}
Comparing this with \eqref{gauss}, we obtain
\begin{align}
\log J &= \frac{\hat{N}^2}2 \log 2\pi - \frac{\hat{N}^2}2 \log \hat{N} 
+ \frac{3}4 \hat{N}^2 + O(\hat{N}), \nonumber \\
J=& \left( \frac{2\pi e^{\frac{3}2}}{\hat{N}} \right)^{\frac{\hat{N}^2}2}.  
\label{J}
\end{align}
{}From \eqref{C} and \eqref{J}, we finally find
\begin{align}
C=\left(\frac{\hN^3}{(2\pi)^3 e^{\frac{3}{2}} g_m^2}\right)^{\frac{\hN^2}{2}}
e^{-\pi i\tau_0 \hN^2/N}. 
\label{Cfinal}
\end{align}

\section{Derivation of the Veneziano-Yankielowicz superpotential}
\setcounter{equation}{0}
Now we make a connection with the $\cN=1$ gauge theory 
coupled to a chiral superfield in the adjoint representation. 
In \eqref{original} we take $m_2=m_3=\Lambda_0 \gg m_1=g_1$. 
Then at a scale below $\Lambda_0$, 
$\Phi_2$ and $\Phi_3$ are decoupled and the system is described 
by the $\cN=1$ gauge theory coupled to the single chiral multiplet 
$\Phi_1$. 
{}From the point of view of this theory, $\Lambda_0$ can be regarded as 
the cutoff and $\tau_0$ as the bare gauge 
coupling there. 
For simplicity, we consider the case where 
the superpotential $W(\Phi_1)$ for $\Phi_1$ is Gaussian: 
$W(\Phi_1)=m_1\Phi_1^2/2$. The general case is considered in Appendix \ref{brokengauge}. 
Then the prepotential in this $\cN=1$ theory should be given 
by $F_{\cN=1^*}$ in \eqref{N=4F} with $m_2=m_3=\Lambda_0$. 
Here we emphasize that the measure factor $C$ is common in the entire range
of $m_i$'s so that we can use the above obtained value \eqref{Cfinal} 
also in the $\cN=1$ limit, where $m_2=m_3 \gg m_1$.
We thus find that the partition function of this matrix model 
is given as 
\begin{align}
e^{-\frac{\hN^2}{g_m^2}F_{\cN=1}}=
Z_{\cN=1}
=&C \int d \Phi_1 d \Phi_2 d \Phi_3 
  e^{-S_{\cN=1^*}(m_2=m_3=\Lambda_0)} \nonumber \\
=&C \left(\frac{2\pi g_m}{\hat{N} \Lambda_0}\right)^{\hat{N}^2} 
 \int  d \Phi_1  e^{-S_{\cN=1}}\nonumber \\
 =&\left( \frac{\hat{N}}{2\pi e^{\frac{3}2}\Lambda_0^2} \right)^{\frac{\hat{N}^2}2}e^{-\pi i\tau_0 \hN^2/N}
 \int  d \Phi_1  e^{-S_{\cN=1}},   
 \label{derivation}
\end{align}
where 
\begin{align}
S_{\cN=1}=\frac{\hN}{g_m}\Tr\frac{m_1}{2}\Phi_1^2,
\end{align}
and we have used the fact that the interaction term can be neglected 
and the $\Phi_2$ and $\Phi_3$ 
integrations are reduced to Gaussian when $\Lambda_0$ is sufficiently large. 
Performing the last integration and identifying $g_m$ with $S$, we obtain
\begin{align}
F_{\cN=1}=&\frac{g_m^2}{2}\left(\frac{2\pi i\tau_0}{N}+\log \frac{e^{\frac{3}{2}}\Lambda _0^2 m_1}{g_m}\right)\nonumber\\
 =&\frac{S^2}{2}\left(\frac{2\pi i\tau_0}{N}+\log \frac{e^{\frac{3}{2}}\Lambda _0^3}{S}\right)+\frac{S^2}{2}\log \frac{m_1}{\Lambda_0},
  \label{finitemass}
\end{align}
which exactly agrees with the prepotential of the $\cN=1$ gauge theory 
that yields the VY superpotential plus the one-loop 
contribution from the chiral multiplet to the gauge kinetic term. 

It is instructive to compare the measure in \eqref{derivation} to that 
in \cite{Cachazo:2002ry}. 
There it is shown that if we consider 
the matrix model with an appropriate measure $\mu$
\begin{align}
\int\frac{d\Phi_1}{\mu^{\hN^2}}e^{-S_{\cN=1}},
\label{mu} 
\end{align}
its free energy reproduces the VY superpotential plus 
the matter contribution provided that 
\begin{align}
\frac{\hat{N}\mu^2}{2\pi}=e^{\frac{3}2}\Lambda_0^2. 
\label{CDSWrelation}
\end{align}  
Eq.\eqref{derivation} shows that 
we can derive this relation directly 
by fixing the measure in the matrix model so that 
\eqref{requirement} will be satisfied.
The extra factor $e^{-\pi i\tau_0\hN^2/N}$ is nothing but the 
contribution from the tree gauge kinetic term, which is again
consistent with the result in \cite{Cachazo:2002ry}. 

By construction, it is evident that we can also obtain 
the VY superpotential in the pure $\cN=1$ gauge theory
by setting $m_1=m_2=m_3=\Lambda_0$. In fact, if we set $m_1=\Lambda_0$ 
in \eqref{finitemass}, we obtain
\begin{align}
F_{\cN=1}^{pure}=\frac{S^2}{2} \log \frac{e^{\frac{3}{2}}\Lambda^3}{S},
  \label{pureYM}
\end{align}
where we have used
\begin{align} 
2\pi i\tau_0 =3N\log\left(\frac{\Lambda}{\Lambda_0}\right).
 \label{tauren}
\end{align}

If we introduce a non-trivial potential to $\Phi _1$, the gauge
symmetry can be broken, and the prepotential becomes a function
of several $S_i$'s. Even in such case, the VY superpotential 
is correctly reproduced by the measure \eqref{Cfinal}. 
In Appendix \ref{brokengauge} we clarify this point
when the $U(N)$ gauge group is broken to $U(N_1)\times U(N_2)$.  

\section{Conclusions}
\setcounter{equation}{0}
Let us summarize the meaning of our results: 
we start from the supersymmetric gauge theory coupled 
to the three chiral multiplets $\Phi_i$ ($i=1\sim 3$) 
with a generic potential as in \eqref{potential} for $\Phi_1$. 
Then the Schwinger-Dyson approach in \cite{Cachazo:2002ry}, or 
the direct map in \cite{Kawai:2003yf} tells us that 
the prepotential $\cF$ in this theory 
and the free energy in the corresponding matrix model $F_m$ 
satisfy
\begin{align}
\frac{\der\cF}{\der g_k}=\frac{\der F_m}{\der g_k},
\end{align}
as a function of $S$ and $g_m$ respectively. Similarly, 
we further find that 
\begin{align}
\frac{\der\cF}{\der m_i}=\frac{\der F_m}{\der m_i},
\end{align}
where $m_i$ ($i=2,3$) is the mass for the chiral multiplet $\Phi_i$. 
Thus $\cF$ and $F_m$ are 
equivalent up to a function independent of $m_i$'s as well as $g_k$'s, 
which is nothing but the contribution from $C$. 
Therefore, if we 
adjust the origin of $F_m$ 
so that $F_m$ and $\cF$ will coincide 
at an appropriate point in the parameter space, 
they will become entirely equivalent. 
We have chosen the $\cN=4$ SYM ($g_k, m_i \rightarrow 0$) 
as such a point, 
where there are no quantum corrections to the holomorphic part 
of the effective Lagrangian. Then in the $\cN=1$ theory 
($m_2=m_3=\Lambda_0$), $F_m$ correctly reproduces 
without any other inputs the VY superpotential 
that is independent of $g_k$'s, as expected.    

Finally we clarify the reason why the matrix model has 
information on the $g_k$-independent part  
in its measure from the point of view of the large-$N$ reduction 
\cite{Kawai:2003yf}. 
{}From the map constructed in \cite{Kawai:2003yf}, we obtain 
the matrix model equivalent to the $\cN=1$ gauge theory 
as far as the holomorphic part is concerned:  
\begin{align}
\exp\left(-\frac{\hN^2}{g_m^2}F_{\cN=1}\right) 
 =  \int d\hPhi \int d\hV 
       \exp\left(-\mmfac\{ 2\pi i\tau_0 \Tr(\hW^{\al}\hW_{\al})
                          +\Tr(W(\hPhi))\}\right),  
\label{supermmfreeE} 
\end{align}
where the hat denotes the large-$N$ reduction of the corresponding 
field in the original $\cN=1$ gauge theory, and 
$F_m$ is the free energy of this model. 
We note here that once we concentrate on the holomorphic part 
of the free energy and drop the kinetic term for $\hPhi$, 
$\hV$ and $\hPhi$ become decoupled from each other,
and the integration over $\hV$ 
can be performed independently 
leaving an overall measure for $\hPhi$.\footnote
{However, this decoupling becomes subtle when there is 
an ultraviolet divergence. See \cite{Kawai:2003yf}.} 
In viewing this, we find that 
the matrix model (\ref{DV action}) is obtained after integrating out the vector 
multiplet, 
and as a consequence, 
if the measure in the matrix model can be determined correctly, 
it should have information on the gauge dynamics. 
In fact, our result shows that this is indeed the case: 
for example, the matrix model reproduces the Veneziano-Yankielowicz 
superpotential, which contains the dynamical scale $\Lambda$ 
for the gauge field. 
The advantage of this standpoint is that 
we can make a direct connection 
between the gauge theory and the matrix model
even for rather complicated multi-matrix models 
such as \eqref{1^*}.

\begin{center} \begin{large}
Acknowledgments
\end{large} \end{center}
This work was supported in part by the Grants-in-Aid for Scientific Research of the Ministry of Education, Culture, Sports, Science and Technology of Japan.
Part of this work has been done while K.Y. was a guest at YITP (Kyoto). 
He would like to thank T.Kugo and F.Ninomiya for their warm hospitality. 
A financial help from INFN (Italy) is gratefully acknowledged.
The work of T.K. was supported in part by the Special Postdoctoral 
Researchers Program. The work of T.M. was supported in part by a Grant-in-Aid for the 21st Century COE "Center for Diversity and Universality in Physics".

\appendix

\section{Derivation of the Veneziano-Yankielowicz superpotential 
in the case of broken gauge symmetry}
\setcounter{equation}{0}
\label{brokengauge}

In this appendix we show how the VY superpotential is derived 
from the matrix model,
when the gauge symmetry is spontaneously broken. 
For simplicity, we concentrate on the case where 
the $U(N)$ gauge symmetry is broken to $U(N_1)\times U(N_2)$ 
($N=N_1+N_2$).
 
We consider the $\hN\times \hN$ matrix model with a cubic potential 
\begin{align}
W(\Phi)=a\left(\frac{1}3 \Phi^3-v^2 \Phi \right),~~~
W'(\Phi)=a(\Phi-v)(\Phi+v),  
\label{cubic}
\end{align}
then the partition function under the measure $\eqref{Cfinal}$ 
is given as in \eqref{derivation} by 
\begin{align}
Z=\frac{1}{J_{\hat{N}}}\left(\frac{1}{\Lambda_0^2}\right)^{\frac{\hat{N}^2}2} e^{-\frac{\pi i \tau_0 }{N} \hat{N}^2} \int d\Phi~ e^{-\frac{\hat{N}}{g_m} \tr W(\Phi) },
\end{align}
where $J_{\hat{N}}$ is defined as in \eqref{J}: 
\begin{align}
J_{\hat{N}}=\left( \frac{2\pi e^{\frac{3}2}}{\hat{N}} \right)^{\frac{\hat{N}^2}2}.
\end{align}
Diagonalizing $\Phi$, we obtain 
\begin{align}
Z&=\left(\frac{1}{\Lambda_0^2}\right)^{\frac{\hat{N}^2}2} e^{-\frac{\pi i \tau_0 }{N} \hat{N}^2}
\int \prod_i d \phi_i \prod_{i< j} (\phi_i-\phi_j)^2 e^{-S'}, 
 \nonumber \\ 
S'&=\frac{\hat{N}}{g_m} a \sum_i \left(\frac{1}{3}\phi_i^3-v^2 \phi_i 
 \right).
\end{align} 
We choose a vacuum where, among $\hN$ eigenvalues, 
$\hN_1$ of them lie around the one classical minimum $v$, 
while the rest $\hN_2$ around the other $-v$. 
Then we consider the fluctuations around it: 
\begin{align}
S'= \frac{\hat{N}}{g_m}\sum_{i=1}^{\hat{N}_1}
\left( \frac{2av}{2} p_i^2+\frac{a}{3}p_i^3 \right)
+\frac{\hat{N}}{g_m}\sum_{i=1}^{\hat{N}_2}
\left( -\frac{2av}{2} q_i^2+\frac{a}{3}q_i^3 \right), 
\label{fluc}
\end{align}
where $p_i$ and $q_i$ are fluctuations around $\phi_i=v$ and 
$\phi_i=-v$ respectively, and we have dropped the constant term. 
Then the partition function becomes 
\begin{align}
Z=& \frac{{}_{\hat{N}} C_{\hat{N}_1}}{\Lambda_0^{\hat{N}^2}}e^{-\frac{\pi i \tau_0 }{N} \hat{N}^2}
\int \prod_{i=1}^{\hat{N}_1} d p_i \prod_{i=1}^{\hat{N}_2}  d q_i \nonumber \\
& \times\prod_{1 \le i < j \le \hat{N}_1} (p_i-p_j)^2  \prod_{1\le i < j \le \hat{N}_2} (q_i-q_j)^2  \prod_{i=1}^{\hat{N}_1} \prod_{j=1}^{\hat{N}_2}(2v+p_i-q_j)^2
e^{-S'},
\label{Zpq}
\end{align}
where ${}_{\hat{N}} C_{\hat{N}_1}$ is the number of the ways 
of choosing $\hN_1$ eigenvalues around $v$. 
We now take the limit 
$2v \rightarrow \Lambda_0 \gg 1$, 
in which the theory becomes the pure $\cN=1$ SYM 
with $U(N_1)\times U(N_2)$ gauge group. 
In this limit $p_i$'s and $q_i$'s are decoupled from each other 
as seen from \eqref{Zpq}, and the cubic terms 
in \eqref{fluc} can be neglected.  
Although in the resulting action $q_i$'s have a negative mass squared, 
we can take an appropriate contour to make the integral convergent 
as usual in the Dijkgraaf-Vafa theory.\footnote
{This is naturally justified 
in \cite{Kawai:2003yf,Dijkgraaf:2003xk} by considering 
a supermatrix model.} 
We thus obtain 
\begin{align}
Z=& \frac{{}_{\hat{N}} C_{\hat{N}_1}}{\Lambda_0^{\hat{N}^2}}e^{-\frac{\pi i \tau_0 }{N} \hat{N}^2}
\int \prod_{i=1}^{\hat{N}_1} d p_i \prod_{i=1}^{\hat{N}_2} d q_i \prod_{1 \le i < j \le \hat{N}_1} (p_i-p_j)^2 \prod_{1\le i < j \le \hat{N}_2} (q_i-q_j)^2 \nonumber \\ 
&\times \prod_{i=1}^{\hat{N}_1} \prod_{j=1}^{\hat{N}_2}
(\Lambda_0+p_i-q_j)^2
\exp\left(-\frac{\hat{N}}{g_m} 
\sum_i\frac{\Lambda_0a}{2}p_i^2  
-\frac{\hat{N}}{g_m} 
\sum_i\frac{\Lambda_0a}{2}q_i^2 \right) \nonumber \\
=& \frac{{}_{\hat{N}} C_{\hat{N}_1}}{\Lambda_0^{\hat{N}^2}}
\frac{\Lambda_0^{2 \hat{N}_1 \hat{N}_2}}
     {J_{\hat{N}_1} J_{\hat{N}_2}}
     e^{-\frac{\pi i \tau_0 }{N} \hat{N}^2}
\left(J_{\hat{N}_1}\int 
\prod_{i=1}^{\hat{N}_1} d p_i  
\prod_{1 \le i < j \le \hat{N}_1} (p_i-p_j)^2 
e^{-\frac{\hat{N}}{g_m} 
\sum_i\frac{\Lambda_0 a}{2}p_i^2}\right) \nonumber \\
& \times\left(J_{\hat{N}_2} \int  
\prod_{i=1}^{\hat{N}_2} d q_i 
\prod_{1\le i < j \le \hat{N}_2} (q_i-q_j)^2 
e^{-\frac{\hat{N}}{g_m} 
\sum_i\frac{\Lambda_0 a}{2}q_i^2}\right) \nonumber \\
=&\frac{{}_{\hat{N}} C_{\hat{N}_1}}{\Lambda_0^{\hat{N}^2}}
\frac{\Lambda_0^{2 \hat{N}_1 \hat{N}_2}}
{J_{\hat{N}_1} J_{\hat{N}_2}}
e^{-\frac{\pi i \tau_0 }{N} \hat{N}^2}
\left( \frac{2\pi g_m}{\hat{N} \Lambda_0a} \right)
^{\frac{\hat{N}_1^2}2}
\left( \frac{2\pi g_m}{\hat{N} \Lambda_0a} \right)
^{\frac{\hat{N}_2^2}2} \nonumber \\
=&{}_{\hat{N}} C_{\hat{N}_1}e^{-\frac{\pi i \tau_0 }{N} \hat{N}^2}
\left( \frac{ g_m \hat{N}_1}{\hat{N} e^{\frac{3}2} \Lambda_0^3a} 
\right)^{\frac{\hat{N}_1^2}2}
\left( \frac{ g_m \hat{N}_2}{\hat{N} e^{\frac{3}2} \Lambda_0^3a} 
\right)^{\frac{\hat{N}_2^2}2},
\end{align}
where we have used the formula \eqref{Jacobian} in a reverse manner such as 
\begin{align*}
J_{\hN_1}\int \prod_{i=1}^{\hN_1} d p_i  \prod_{1 \le i < j \le \hN_1} (p_i-p_j)^2 e^{-\frac{\hN}{g_m} \sum_i\frac{m}{2}p_i^2} 
= \int d^{\hN_1^2} \tilde\Phi 
e^{-\frac{\hN}{g_m}\tr \frac{m}2 \tilde\Phi^2}
=\left( \frac{2\pi g_m}{\hN m} \right)^{\frac{\hat{N}_1^2}2}. 
\end{align*}
Therefore, we obtain the free energy in the large-$\hN$ limit as 
\begin{align}
F_m=
\frac{\pi i \tau_0 }{N} g_m^2+ \frac{g_m^2 \hat{N}_1^2}{2\hat{N}^2} \log \left( \frac{\hat{N} e^{\frac{3}2} \Lambda_0^3a}{ g_m \hat{N}_1} \right)
+\frac{g_m^2 \hat{N}_2^2}{2\hat{N}^2} \log \left( \frac{\hat{N} e^{\frac{3}2} \Lambda_0^3a}{ g_m \hat{N}_2} \right).
\end{align}
Substituting $S_i$ for $g_m \hat{N}_i/ \hat{N}$, 
we finally obtain 
\begin{align}
F_m=
\frac{\pi i \tau_0 }{N} \left(S_1+S_2\right)^2
+ \frac{S_1^2}2 \log \left( \frac{ e^{\frac{3}2} \Lambda_0^3a}{ S_1} \right)
+ \frac{S_2^2}2 \log \left( \frac{ e^{\frac{3}2} \Lambda_0^3a}{ S_2} \right). 
\end{align}
According to the formula for the effective superpotential  
in the case of broken gauge symmetry \cite{Cachazo:2002ry}
\begin{align}
W^{\text{eff}}=\sum_iN_i\frac{\der \cF}{\der S_i},
\end{align}
we find that under the identification $\cF=F_m$, 
$F_m$ exactly reproduces the VY superpotential 
when $U(N)$ gauge symmetry is broken to $U(N_1)\times U(N_2)$, 
where $S_i=\frac{1}{64\pi^2}\tr_{U(N_i)} W^{\al}W_{\al}$.


\begin{thebibliography}{99}

\bibitem{Kawai:1982nm}
T.~Eguchi and H.~Kawai,
``Reduction Of Dynamical Degrees Of Freedom In The Large N Gauge Theory,''
Phys.\ Rev.\ Lett.\  {\bf 48}, 1063 (1982); \\
G.~Parisi,
``A Simple Expression For Planar Field Theories,''
Phys.\ Lett.\ B {\bf 112}, 463 (1982); \\
D.~J.~Gross and Y.~Kitazawa,
``A Quenched Momentum Prescription For Large N Theories,''
Nucl.\ Phys.\ B {\bf 206}, 440 (1982); \\
G.~Bhanot, U.~M.~Heller and H.~Neuberger,
``The Quenched Eguchi-Kawai Model,''
Phys.\ Lett.\ B {\bf 113}, 47 (1982); \\
S.~R.~Das and S.~R.~Wadia,
``Translation Invariance And A Reduced Model For Summing Planar Diagrams In QCD,''
Phys.\ Lett.\ B {\bf 117}, 228 (1982); \\
A.~Gonzalez-Arroyo and M.~Okawa,
``The Twisted Eguchi-Kawai Model: A Reduced Model For Large N Lattice Gauge Theory,''
Phys.\ Rev.\ D {\bf 27}, 2397 (1983).
    

\bibitem{Ishibashi:1996xs}
N.~Ishibashi, H.~Kawai, Y.~Kitazawa and A.~Tsuchiya,
``A large-N reduced model as superstring,''
Nucl.\ Phys.\ B {\bf 498}, 467 (1997)
[arXiv:hep-th/9612115].


\bibitem{Dijkgraaf:2002dh}
R.~Dijkgraaf and C.~Vafa,
``A perturbative window into non-perturbative physics,''
arXiv:hep-th/0208048.


\bibitem{Dijkgraaf:2002xd}
R.~Dijkgraaf, M.~T.~Grisaru, C.~S.~Lam, C.~Vafa and D.~Zanon,
``Perturbative computation of glueball superpotentials,''
Phys.\ Lett.\ B {\bf 573}, 138 (2003)
[arXiv:hep-th/0211017].


\bibitem{Cachazo:2002ry}
F.~Cachazo, M.~R.~Douglas, N.~Seiberg and E.~Witten,
``Chiral rings and anomalies in supersymmetric gauge theory,''
JHEP {\bf 0212}, 071 (2002)
[arXiv:hep-th/0211170].


\bibitem{Konishi:1983hf}
K.~Konishi,
``Anomalous Supersymmetry Transformation Of Some Composite Operators In Sqcd,'' Phys.\ Lett.\ B {\bf 135}, 439 (1984); \\
K.~i.~Konishi and K.~i.~Shizuya,
``Functional Integral Approach To Chiral Anomalies In Supersymmetric Gauge Theories,''
Nuovo Cim.\ A {\bf 90}, 111 (1985).


\bibitem{Ooguri:2002gx}
H.~Ooguri and C.~Vafa,
``Worldsheet derivation of a large N duality,''
Nucl.\ Phys.\ B {\bf 641}, 3 (2002)
[arXiv:hep-th/0205297].


\bibitem{Veneziano:1982ah}
G.~Veneziano and S.~Yankielowicz,
``An Effective Lagrangian For The Pure N=1 Supersymmetric Yang-Mills 
Theory,''
Phys.\ Lett.\ B {\bf 113}, 231 (1982).


\bibitem{Kawai:2003yf}
H.~Kawai, T.~Kuroki and T.~Morita,
``Dijkgraaf-Vafa theory as large-N reduction,''
Nucl.\ Phys.\ B {\bf 664}, 185 (2003)
[arXiv:hep-th/0303210]; \\
H.~Kawai, T.~Kuroki and T.~Morita,
``Supersymmetric large-N reduced model with multiple matter,''
Nucl.\ Phys.\ B {\bf 683}, 27 (2004)
[arXiv:hep-th/0312026].

\bibitem{Buchbinder:2004bu}
I.~L.~Buchbinder,
``Progress in study of N = 4 SYM effective action,''
arXiv:hep-th/0402089; \\
I.~L.~Buchbinder and S.~M.~Kuzenko,
``Comments on the background field method in harmonic superspace: 
Non-holomorphic corrections in N = 4 SYM,''
Mod.\ Phys.\ Lett.\ A {\bf 13}, 1623 (1998)
[arXiv:hep-th/9804168]; \\
M.~Dine and N.~Seiberg,
``Comments on higher derivative operators in some SUSY field theories,''
Phys.\ Lett.\ B {\bf 409}, 239 (1997)
[arXiv:hep-th/9705057]; \\
See also N.~Dorey,
``An elliptic superpotential for softly broken N = 4 supersymmetric  Yang-Mills theory,''
JHEP {\bf 9907}, 021 (1999)
[arXiv:hep-th/9906011].


\bibitem{Gripaios:2003kt}
B.~M.~Gripaios and J.~F.~Wheater,
``Veneziano-Yankielowicz superpotential terms in N = 1 SUSY gauge  theories,''
Phys.\ Lett.\ B {\bf 587}, 150 (2004)
[arXiv:hep-th/0307176].


\bibitem{Arnone:2004ek}
S.~Arnone, F.~Guerrieri and K.~Yoshida,
``N = 1* model and glueball superpotential from renormalization group improved perturbation theory,''
JHEP {\bf 0405}, 031 (2004)
[arXiv:hep-th/0402035].


\bibitem{Dijkgraaf:2003xk}
R.~Dijkgraaf and C.~Vafa,
``N = 1 supersymmetry, deconstruction, and bosonic gauge theories,''
arXiv:hep-th/0302011.



\end{thebibliography}
\end{document}